\documentclass[twocolumn,showpacs,preprintnumbers,amsmath,amssymb,10pt,prd]{revtex4}

\usepackage{fancyhdr}
\usepackage{amsmath,amsfonts,amssymb}

\usepackage{color}
\definecolor{red}{rgb}{1,0,0}
\definecolor{gre}{rgb}{0,0.6,0}
\definecolor{blu}{rgb}{0,0,1}

\usepackage{graphicx} 
\usepackage{mathrsfs}
\DeclareGraphicsExtensions{.jpg, .JPG , .png , .pdf, .gif}

\newcommand{\lb}[1]{\label{#1}}

\newcommand{\be}{\begin{equation}}
\newcommand{\ee}{\end{equation}}
\newcommand{\ba}{\begin{eqnarray}}
\newcommand{\ea}{\end{eqnarray}}
\newcommand{\eq}[1]{Eq. (\ref{#1})}
\newcommand{\eqs}[2]{Eqs. (\ref{#1}) - (\ref{#2})}
\newcommand{\pd}[2]{\frac{\partial#1}{\partial#2}}

\newcommand{\sign}{\mathrm{sign}}

\newcommand{\h}{\mathcal{H}}

\begin{document}
\title{Restrictions on curved cosmologies in modified gravity from metric considerations}

\author{Linda Linsefors}%
 \email{linsefors@lpsc.in2p3.fr}
\affiliation{%
Laboratoire de Physique Subatomique et de Cosmologie, Universite Grenoble-Alpes, CNRS-IN2P3\\
53, avenue des Martyrs, 38026 Grenoble cedex, France
}%

\author{Aurelien Barrau}%
 \email{Aurelien.Barrau@cern.ch}
\affiliation{%
Laboratoire de Physique Subatomique et de Cosmologie, Universite Grenoble-Alpes, CNRS-IN2P3\\
53,avenue des Martyrs, 38026 Grenoble cedex, France
}%

\date{\today}

\begin{abstract}
This study uses very simple symmetry and consistency considerations to put constraints on possible Friedmann equations for modified gravity models in curved spaces. As an example, it is applied to loop quantum cosmology.
\end{abstract}


\maketitle

\section{Introduction}

From the mathematical perspective, the equations governing the background cosmological evolution can be seen as is a symmetry reduced version of the gravity field equations. As well as being successful in describing the evolution of the universe, cosmology can be seen as an interesting testing ground for new theories of gravity, in particular motivated for being effective models (or low-energy limits) of quantum gravity.\\

In this article, we will study a general class of modified cosmologies that will be defined by a number of assumptions. We will find how these theories are constrained by the coordinate freedom that is fundamentally encoded in the metric, whatever the considered theory. Beyond constraints on the modified Friedmann equations, an interesting result will be to show how the Hamiltonian can also be constrained for consistency reasons.

Our study is rooted in the symmetries of de Sitter and Minkowski spaces. Intuitively speaking, the idea is to consider a de Sitter phase and use its maximal symmetry.\\

As a fruitful example, the conclusions previously derived will be applied to loop quantum cosmology (LQC), see \cite{lqc_review} for general introductions. In itself, LQC is a symmetry reduced version of loop quantum gravity, see \cite{lqg_review} for introductory reviews. Basically, in LQC, the Big Bang is generically replaced by a Big Bounce and sharply peaked states have been shown to be well described by known effective equations.

\section{FLRW metric}

The FLRW metric reads as
\be\lb{flrw}
ds^2=-dt^2+a(t)^2\left(\frac{dr^2}{1-k\,r^2}+r^2d\Omega^2\right).
\ee
This is the most general homogenous and isotropic metric one can write down. More precisely, this is the interval written in a coordinate system where the symmetries of the Universe are clearly manifest. The only way to preserve the homogeneity and isotropy of space and yet incorporate time evolution is to allow the curvature scale, characterized by $a$, to be time-dependent. At this stage, only symmetries are involved and nothing is assumed about the details of the considered gravitational theory. In this expression, $k$ is a constant and $a(t)$ is the scale factor. The evolution of $a(t)$ is determined by Einstein's equations or, alternatively, by some modified gravity or modified cosmology theory.\\

In general, there are two possible coordinate transformation witch leave the FLRW formalism invariant. The first one is a re-scaling of the radial coordinate by a constant $b>0$. Such a transformation affects both $a$ and $k$, but keeps the FLRW expression unchanged:
\ba\lb{trans1}
&&r'= r/b, \\
&&a'(t)= b\, a(t),\\ 
&&k'= b^2\, k. \lb{trans2}
\ea

The other possibility is a time translation, which is of no interest in this study.

It is common in the literature to fix this coordinate freedom by choosing $k\pm1$ whenever $k\neq0$.
We will not do so in this article.

\section{(Modified) Friedman Equation}

Classically, the evolution of $a(t)$ is given by the first Friedmann equation,
\be\lb{clFried}
H^2=-\frac{k}{a^2}+\frac{\kappa}{3}\rho + \frac{\Lambda}{3},
\ee
where $H:=\frac{\dot{a}}{a}$ is the Hubble parameter, $\kappa=8\pi G$, $\rho$ is the matter energy density, and $\Lambda$ is the cosmological constant. 
This equation is correct for any type of homogenous matter. By homogenous we mean that $\rho$ is constant over space-like slices defined by a constant value to the time variable $t$.

The first Friedmann equation is directly derived from general relativity (GR) field equations, or alternatively from the Hamiltionan constraint. A modified theory of gravity (that may or may not come out from some version of quantum gravity) will most probably give rise to a modified Friedmann equation.\\

It should be noticed that the Friedmann Eq. (\ref{clFried}) is invariant under the rescalings given by \eqs{trans1}{trans2}. Any modified Friedmann equation must have this property. Otherwise, the theory would be inconsistent, or alternatively \eq{flrw} would not describe a metric. 

The first Friedmann equation (that we are interested in for this study) is a reformulation of the Hamiltionan constraint, this is why it only involves first order derivatives. We assume that this will also be the case for the equation of motion of $a$ in the modified cosmology considered here.
Since we are restricted to first order derivatives of $a$ in Eq (\ref{clFried}), there are only three independent gravitational variables as far as this specific equation in concerned: $a$, $\dot{a}$ and $k$. From these, we can construct two independent gravitational quantities that are invariant under \eqs{trans1}{trans2}: $H$ and $\frac{k}{a^2}$. The equation of motion for $H^2$ can in principle always be solved and the result has to be a function of $\frac{k}{a^2}$ and matter variables.

\subsection{Main assumptions and their consequences}

The assumptions so far for the modified cosmology or modified gravity theory considered are:

\begin{itemize}
\item[\textbf{1.}] If the universe starts out homogenous and isotropic, it remains homogenous and isotropic. This is certainly not true at all scales as any consistent theory should lead to a growth of inhomogeneities. But this is very reasonable at the background order.
\item[\textbf{2.}] The theory allows for a metric interpretation, {\it i.e.} all physical equations must be invariant under metric coordinate transformations.
\item[\textbf{3.}] Given the metric, \eq{flrw}, the equation of motion for the scale factor $a(t)$ is given by the first Friedmann equation or its analogous in the modified theory considered, which is first order in the time derivative of $a(t)$.
\item[\textbf{4.}] There are no hidden gravitational degrees of freedom apart from the metric.
\end{itemize}

Any theory of modified gravity or modified cosmology that fulfills the above assumptions will have a (modified) Friedmann equation of the form
\be\lb{modFried1}
H^2=\tilde{f}\left(\frac{k}{a^2},\text{matter}\right),
\ee
where $\tilde{f}$ is a function of $\frac{k}{a^2}$ and of any set of homogenous coordinate-independent matter variables. This is grounded in the symmetries.\\

It can be noticed that in the flat case, $k=0$, the modified Friedmann equation is not allowed to depend explicitly on $a$. This is of course true in GR.

\subsection{Additional assumptions}

It is now  necessary add two more assumptions to go ahead in the study. 

\begin{itemize}
\item[\textbf{5.}] The total energy density is the only matter variable that enters the first modified Friedmann equation.
\end{itemize}
By combining the above assumption with \eq{modFried1}, one gets
\be\lb{modFried2}
H^2=f\left(\frac{k}{a^2},\rho\right),
\ee
where $f$ is a function of $\frac{k}{a^2}$ and $\rho$.

\begin{itemize}
\item[\textbf{6.}] Given an arbitrary constant $\rho_1$ such that $f\left(\frac{k}{a^2},\rho_1\right)\geq0$, the theory allows $\rho=\rho_1$ for at least a non-vanishing amount of time.
\end{itemize}
A situation with a constant energy density could for example be realized by a scalar field temporarily trapped in a false vacuum, or by a vacuum quantum-fluctuations domination stage. It is important to stress that we don't need this specific stage to have been explicitly realized in the history of the Universe, we just need the theory to be able to account for such a stage. This is obviously the case for GR and for all the most discussed theories beyond GR.\\

In the analysis performed so far, the possibility of a cosmological constant and/or dark energy has not been left out. If the acceleration of the universe is due to some exotic matter content (dark energy), then this will be included in $\rho$. If, on the other hand, the acceleration of the universe is due to a true cosmological constant $\Lambda$, this will be included directly in the function $f$ by the relation $\frac{\Lambda}{3}=f(0,0)$.

\subsection{de Sitter / Minkovski space-time}

Let us choose a situation where $k=0$ and $\rho=\rho_1$ such that $f(0,\rho_1)\geq0$ for some time. Then we have: 
\be \lb{con}
H^2=f(0,\rho_1)=\text{constant},
\ee
for a non-vanishing amount of time.
The above equation together with the FLRW metric, \eq{flrw}, describes exactly the de Sitter space-time for $f(0,\rho_1)>0$, and Minkowski space-time for $f(0,\rho_1)=0$.\\ 

By choosing a specific situation where $\rho$ is constant in time, we get an extra symmetry of the system. In the general case, \eqs{trans1}{trans2} are the only coordinate transformations that preserve the FLRW formulation. However, due to the time symmetry of Minkowski and de Sitter space-times, more coordinate transformations are available still within the FLRW metric formulation.

Using the coordinate transformation described in Appendix A, one finds that 
\be \lb{cur}
H^2=-\frac{k}{a^2}+f(0,\rho_1) \ ,\quad \forall k\leq a^2f(0,\rho_1),
\ee
describes exactly the same space-time as \eq{con}. Therefore, if \eq{con} is correct then \eq{cur} must be correct too.

For any theory of modified gravity or cosmology that fulfill Assumptions (1) - (6), the modified Friedmann equation must therefore be of the form:
\be\lb{modFried3}
H^2=-\frac{k}{a^2}+f_0(\rho),
\ee
where $f_0$ is a function of $\rho$ related to previous expressions by $f_0(\rho)=f(0,\rho)$.

\subsection{Preliminary conclusion}

For a wide large class of modified cosmology models, it was shown that the modified Friedman equation for curved (i.e. $k\neq$0) FLRW space-times, can be immediately derived from the modified Friedman equation for flat a (i.e. $k=$0) FLRW space-time by \eq{modFried3}. This basically relies on the symmetries and should be considered as a ground before going ahead.

\section{Hamiltonian} \lb{secHam}
In this section, the Hamiltonian that leads to \eq{modFried3} will be derived, as far as it is possible without assuming an explicit expression for $f_0(\rho)$. We somehow follow the reverse path when compared to the one usually considered: we begin by finding the Friedmann equation where the constraints can easily be put and the physical meaning of all terms is clear and use it to infer the Hamiltonian. \\

To avoid infinities we consider a finite region of space defined by a fiducial volume ${\cal V}$, given by some fixed region in coordinate space. It follows from the metric that ${\cal V}$ has the volume $V=vV_0$, where $v:=a^3$ and $V_0$ is a constant. We  choose $v$ and $\alpha$ to be the canonical coordinates describing the gravitational degree of freedom. The coordinate $\alpha$ is defined by the Poisson bracket 
\be
\{\alpha,v\}=\frac{1}{V_0}.
\ee
This choice can be made without any loss of generality as it is always possible to change to another pair after the Hamiltonian constraint has been found.

We assume, as usually done, that the matter part of the Hamiltonian is unmodified with respect to the classical case. This defines the total Hamiltonian:
\be\lb{H}
\h_{tot}=\h_G(v,\alpha)+V\rho,
\ee
which could also be seen as a definition of $\rho$.\\

We now derive the expression of the gravitational part of the Hamiltonian $\h_G(v,\alpha)$, using \eqs{modFried3}{H}.

From the Hamiltonian constraint, $\h_{tot}=0$, we get 
\be\lb{rho}
\rho=\frac{-\h_G}{vV_0}.
\ee
We also have:
\be\lb{ho}
H=\frac{\dot{v}}{3v}=\frac{1}{3v} \{v,\h\}=\frac{1}{3vV_0}\pd{(-\h_G)}{\alpha}.
\ee
Combining \eq{rho} and \eq{ho} with \eq{modFried3}, one gets
\be
\pd{(-\h_G)}{\alpha}=\pm 3vV_0 \sqrt{-\frac{k}{v^{2/3}}+ f_0\left(\frac{-\h_G}{vV_0}\right) }.
\ee
Since the RHS of the above equation does not depend explicitly on $\alpha$, one can separate the variables and integrate:
\be \lb{int}
\int d\alpha = \pm \int \frac{d\left(\frac{-\h_G}{vV_0}\right)}{3\sqrt{-\frac{k}{v^{2/3}}+ f_0\left(\frac{-\h_G}{vV_0}\right)  }}
\ee
where $v$ is held constant during the integration. 

When $f_0$ is known, the integration can in principle be performed. Finally, one has to solve for $\h_G$ to obtain the expression for the gravitational part of the Hamiltonian constraint.

\section{Effective LQC}
We now focus on  effective loop quantum cosmology as an example of modified cosmology grounded in quantum gravity consideration.
In LQC, for $k=0$, the Friedmann equation is known to be \cite{lqc_review}:
\be
H^2=\frac{\kappa}{3}\rho\left(1-\frac{\rho}{\rho_c}\right). 
\ee
This is the effective description of the bounce that replaces the Big Bang: the density is bounded from above at the value $\rho_c\sim \rho_{Pl}$ and the Hubble parameter vanishes when this density is reached.
According to the previously given arguments, the Friedmann equation for a general $k$ must be
\be\lb{modFriedLQC}
H^2=-\frac{k}{a^2}+\frac{\kappa}{3}\rho\left(1-\frac{\rho}{\rho_c}\right).
\ee
This is in conflict with earlier results, as it will be discussed in the next section.

\subsection{Hamiltionan}
It is now possible to use results from Section \ref{secHam} to calculate the Hamiltonian that leads to \eq{modFriedLQC}. In this case, 
\be
f_0(\rho)=\frac{\kappa}{3}\rho\left(1-\frac{\rho}{\rho_c}\right).
\ee
Inserting this in \eq{int} gives
\be\lb{ah}
\int d\alpha=
\pm \int \frac{d\left(\frac{-\h_G}{V_0}\right)}{\sqrt{-9kv^{1/3} +3\kappa v\frac{-\h_G}{V_0}\left(1-\frac{1}{3v\rho_c}\frac{-\h_G}{V_0}\right)  }}.
\ee
This can be solved to:
\be\lb{hG}
\h_G=-vV_0\frac{\rho_c}{2}\left(1-\sqrt{1-\frac{12}{\kappa \rho_c}\,\frac{k}{v^{2/3}}}\cos\left(\sqrt{\frac{3\kappa}{\rho_c}}[\alpha-\alpha_1(v)]\right)\right),
\ee
where $\alpha_1$ is an integration `constant'. Since $v$ was kept fixed during the integration, $\alpha_1$ can be any function of $v$. It is easy to check that this Hamiltonian indeed gives the correct modified Friedmann equation.

In the derivation of \eq{hG} no other assumption or ansatz than the one given by the modified Friedmann equation, together with \eq{H}, was assumed. Because of this, \eq{hG} gives all the solutions to $\h_G$, given \eq{modFriedLQC} and \eq{H}.\\

In this study we have chosen to work with the variables $v$ and $\alpha$ for simplicity, and to clarify the dependence upon $\rho_c$ which, together with the coupling constant $\kappa=8\pi G$, is the only parameter entering the dynamics. However, \eq{hG} can be re-expressed using more familiar variables often used in the literature. In the effective formulation, the choice of canonical variables is just a matter of taste. The Hamiltonian can as well be expressed as
\be\lb{hG1}
\h_G=-vV_0\frac{\rho_c}{2}\left(1-\sqrt{1-\frac{12}{\kappa \rho_c}\,\frac{k}{v^{2/3}}}\cos\big(2\lambda[\beta-\beta_1(v)]\big)\right),
\ee
where $\{\beta,v\}=\frac{\gamma\kappa}{2V_0}$, or
\be\lb{hG2}
\h_G=-p^{3/2}V_0\frac{\rho_c}{2}\left(1-\sqrt{1-\frac{12}{\kappa \rho_c}\,\frac{k}{p}}\cos\left(2\frac{\lambda}{\sqrt{p}}[c-c_1(p)]\right)\right),
\ee
where $p=a^2=v^{2/3}$, and $\{c,p\}=\frac{\gamma\kappa}{3V_0}$.

\section{Previous LQC models with $k\neq0$}

Prior to this work, two different curved LQC models were considered for $k>0$ and one for $k<0$. We  briefly review the results in this section. 

As mentioned earlier, usually $k>0$ is referred to as $k=1$ and $k<0$ as $k=-1$, since one often chooses coordinates so that $|k|=1$ using the coordinate freedom of \eqs{trans1}{trans2}.

\subsection{$k>0$, first model: the homology way}
This model was developed independently in both \cite{Szulc:2006ep} and \cite{Ashtekar:2006es}. The effective equations from this Hamiltonian was fist calculated in \cite{Mielczarek:2009kh}, and later in \cite{Corichi:2011pg}.

The effective Hamiltonian in this model is 
\be
\mathcal{H}_{tot}=-V\rho_c\left[\sin^2(\lambda\beta-D)-\sin^2D+\left(1+\gamma^2\right)D^2\right]+ V\rho,
\ee
where $V$ is the total volume of the universe, which makes sens since the Universe is closed, $\{\beta,V\}=\frac{\kappa\gamma}{2}$ and $D=\lambda\left(\frac{2\pi^3}{V}\right)^{1/3}=\lambda\sqrt{\frac{k}{a^2}}$. 

The modified Friedman equation in this model is
\be
H^2=\frac{\kappa}{3}(\rho-\rho_1)\left(1-\frac{\rho-\rho_1}{\rho_c}\right),
\ee
where
\be
\rho_1:=\frac{3}{\kappa}\left[\left(1+\frac{1}{\gamma^2}\right)\frac{k}{a^2}-\frac{1}{\gamma^2\lambda^2}\sin^2\left(\lambda\sqrt{\frac{k}{a^2}}\right)\right].
\ee

\subsection{$k>0$, second model: the connection way}
This model was first suggested in \cite{Ashtekar:2009um}, and further studied in \cite{WilsonEwing:2010rh}.
The effective equations were first derived in \cite{Corichi:2011pg}.

The effective Hamiltonian in this model is 
\be
\mathcal{H}_{tot}=-V\rho_c\left[\left(\sin\lambda\beta-D\right)^2+\gamma^2D^2\right]+V\rho,
\ee
where $V$, $\beta$ and $D$ are the same as in the previous model.

The modified Friedmann equation in this model is
\be
H^2=\left(\frac{\kappa}{3}{\rho}-\frac{k}{a^2}\right)\left(1-\frac{\rho-\rho_2}{\rho_c}\right),
\ee
where
\be
\rho_2=\frac{3}{\kappa}\left[\left(1-\frac{1}{\gamma^2}\right)\frac{k}{a^2}\mp \frac{1}{\gamma}\sqrt{\frac{k}{a^2}\left(\frac{\kappa}{3}\rho-\frac{k},{a^2}\right)}\right]
\ee
and
\be
(\mp)=-\sign\left(\sin\lambda\beta-\lambda\sqrt{\frac{k}{a^2}}\right).
\ee

\subsection{$k<0$}

The effective Hamiltonian in this model, proposed in \cite{Vandersloot:2006ws}, is
\be
\mathcal{H}_{tot}=-\frac{3V_0p^{3/2}}{\kappa\gamma^2\lambda^2}\sin^2\left(\frac{\lambda\,c}{\sqrt{p}}\right)
+\frac{3V_0\sqrt{p}}{\kappa} + V_0p^{3/2}\rho,
\ee
where $\{c,p\}=\frac{\gamma\kappa}{3V_0}$ and $k=-1$. Changing canonical variables, and un-freezing $k$ we get an equivalent expression:
\be
\mathcal{H}_{tot}=-vV_0\rho_c \left[\sin^2\lambda\beta
+\gamma^2\lambda^2\frac{k}{v^{2/3}} \right]
+ vV_0\rho,
\ee
where $\{\beta,v\}=\frac{\gamma\kappa}{2V_0}$.

The modified Friedmann equation in this case is
\be
H^2=\left(\frac{\kappa}{3}\rho-\frac{k}{a^2}\right)\left(1+\gamma^2\lambda^2\frac{k}{a^2}-\frac{\rho}{\rho_c}\right).
\ee

\section{Discussion}

We have shown that under very general and conservative assumptions, for any modified cosmology or modified gravity, the Friedmann equation for curved space can be immediately deduced from the Firedmann equation for a flat space. We have applied this result to effective LQC. In reviewing previous models for effective LQC on curved space, we find that there is a conflict between this work and previous results.\\



The source of this tension might come from the fact the holonomy corrections are in some approaches expected to lead to a deformed algebra of constraints that is known to lead to a non-metric theory. In that sense, the mismatch observed could be taken as an independent indication that holonomy modifications should lead to non-classical space-time structures, or deformed constraint algebras in the canonical formulation.\\

This works anyway suggest to more carefully consider the consistency conditions required for modified cosmologies. Even if there are good theoretical reasons to consider a given Hamiltonian and related modified Friedmann equation, symmetry considerations should not be forgotten if the theory is to be interpreted as a metric theory. \\



\section{Acknowledgments}

We would like to thank Martin Bojowald and Jakub Mielczarek for interesting comments. \\

L.S. is supported by the Labex ENIGMAS.

\appendix

\section{Different FLRW coordinates for de Sitter and Minkowski spaces}
In this Appendix we show by construction that in de Sitter and Minkowski spaces, there is always a coordinate transformation that leads to a FLRW coordinate representation, with the intrinsic curvature of any choice, only limited by $\frac{k}{a^2}\leq\frac{1}{\alpha^2}$, where $\alpha$ is a parameter of the manifold to be defined below.

Intuitively we use the fact that in de Sitter space, curvature is pure gauge.

\subsection{de Sitter space}

\begin{figure}
\includegraphics[width=0.9\columnwidth]{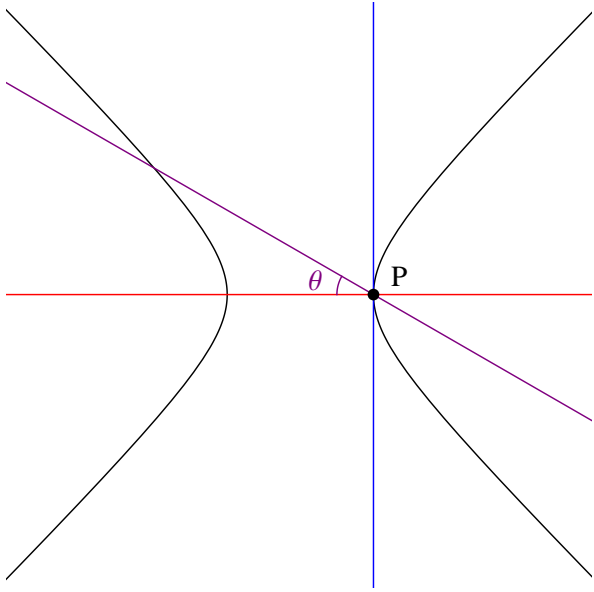}
		\caption{The $x_0,x_1$-plane of Minkowski space. The black lines are the embedding of de Sitter space for R=0. The straight lines represent different FLRW coordinate choices of de Sitter. There is one for each $\theta\in[-\frac{\pi}{2},\frac{\pi}{2}]$.}
		\label{planes}
\end{figure}

A N dimensional de Sitter space can be embedded in a N+1 dimensional Minkowski space. Specifically for N=4, the embedding is given by
\be \lb{dsCord}
-x_0^2+\sum_{i=1}^4x_i^2=\alpha^2,
\ee
and the metric is directly inherited,
\be\lb{sumdx}
ds^2=-dx_0^2+\sum_{i=1}^4dx_i^2,
\ee

where $\alpha$ is defined by \eq{dsCord} and $\alpha\geq0$. Actually, $\alpha$ should be viewed as a property of the manifold rather than one of the coordinates. 

Both the above equations are invariant under a Lorentz transformation of the $N+1$ dimensional Minkowski space. By such a coordinate transformation, any point $P$ in the de Sitter space can be rotated so as to have the coordinates $(x_0,x_1,x_2,x_3,x_4)=(0,\alpha,0,0,0)$.

Given a plane in the larger Minkowski space, 
\be\lb{plane}
b_0 x_0 + b_1 x_1 = c,
\ee
for some constants $b_0$, $b_1$ and $c$ ; the intersection between \eq{dsCord} and \eq{plane} will be a homogenous, 3 dimensional (possibly disconnected) surface. This intersection will be space-like if and only if the plane is intersected by the line $-x_0^2+x_1^2=\alpha^2$ at least once, and light-like if it is tangent to this line.

One can use the intersection with such planes to find FLRW coordinates for the de Sitter space. We start from the ansatz
\ba\lb{ansatz1}
x_0(t,r)&=& \alpha\left[\sinh\left(\frac{t}{\alpha}\right)+\sin(\theta)g(r,t)\right],\qquad\\
x_1(t,r)&=& \alpha\left[\cosh\left(\frac{t}{\alpha}\right)-\cos(\theta)g(r,t)\right],\\
\sqrt{x_2^2+x_3^2+x_4^2}&=& a(t) r,
\ea
where
\be\lb{ansatz2}
g(0,t)=0 \ ,\quad
\theta\in[-\frac{\pi}{2},-\frac{\pi}{2}] ,
\ee
and $t$, $r$ and $a(t)$ are the same as in \eq{flrw}.

The ansatz is chosen so that $t$ is the proper time along $r=0$, and points at constant $t$ belong to a plane at an angle $\theta$ which is defined by Fig. \ref{planes}.

The remaining step is, for every $\theta\in[-\frac{\pi}{2},\frac{\pi}{2}]$, to find $g(t,r)$, $a(t)$ and $k$, so that \eq{dsCord} is fulfilled, and \eq{sumdx} together with \eqs{ansatz1}{ansatz2} yield the FLRW metric.\\

In the case $\sin^2(\theta)\neq\cos^2(\theta)$, it is straightforward to show that the above requirements are fulfilled by
\be
g(t,r)=
\frac{\sin(\theta)\sinh\left(\frac{t}{\alpha}\right)+\cos(\theta)\cosh\left(\frac{t}{\alpha}\right)}{\sin^2(\theta)-\cos^2(\theta)}\left(-1+\sqrt{1-kr^2}\right),
\ee
and
\be
a(t)=\alpha\sqrt{k}\,\frac{\left|\sin(\theta)\sinh\left(\frac{t}{\alpha}\right)+\cos(\theta)\cosh\left(\frac{t}{\alpha}\right)\right|}{\sqrt{\cos^2(\theta)-\sin^2(\theta)}}.
\ee

In the case $\theta=\frac{\pi}{4}$, one has:
\be
g(t,r)=\frac{a_0^2r^2}{\sqrt{2}\, \alpha^2}e^{t/\alpha}
\ ,\quad a(t)=a_0e^{t/\alpha} 
\ ,\quad k=0.
\ee

In the case $\theta=-\frac{\pi}{4}$, one has:
\be
g(t,r)=\frac{a_0^2r^2}{\sqrt{2}\, \alpha^2}e^{-t/\alpha}
\ ,\quad a(t)=a_0e^{-t/\alpha}
\ ,\quad k=0.
\ee

In all the above cases, we find that
\be
H^2=\frac{1}{\alpha^2}-\frac{k}{a^2},
\ee
and 
\be\lb{apFried}
\left.\frac{k}{a^2}\right|_{t=0}=\frac{1}{\alpha^2}\left(1-\tan^2(\theta)\right).
\ee
From the above equation, it is clear that $\frac{k}{a^2}\leq\frac{1}{\alpha^2}$, but other than that, it is a pure coordinate choice. It should be kept in mind that any point can be moved to $t=0$ by a Lorentz transformation of \eq{dsCord}. Therefore, if one considers a FLRW metric with a Friedmann equation on the form \eq{apFried}, one can always move to some other FLRW coordinates with a different value of $\frac{k}{a^2}$ for some given time. Then, \eq{apFried} will still be true for the new coordinates, with the same value of $\alpha$.\\

A special case of this is when $k=0$ in the fist set of coordinates: given a FLRW metric and a Friedmann equation that looks like 
\be
H^2=\text{constant},
\ee
it is always possible to do a coordinate transformation to a system with 
\be
\left(H'\right)^2=\text{constant}-\frac{k'}{\left(a'\right)^2},
\ee
for $\frac{k'}{a'^2}\leq\text{constant}$.

\subsection{Minkowski space}

\begin{figure}
\includegraphics[width=\columnwidth]{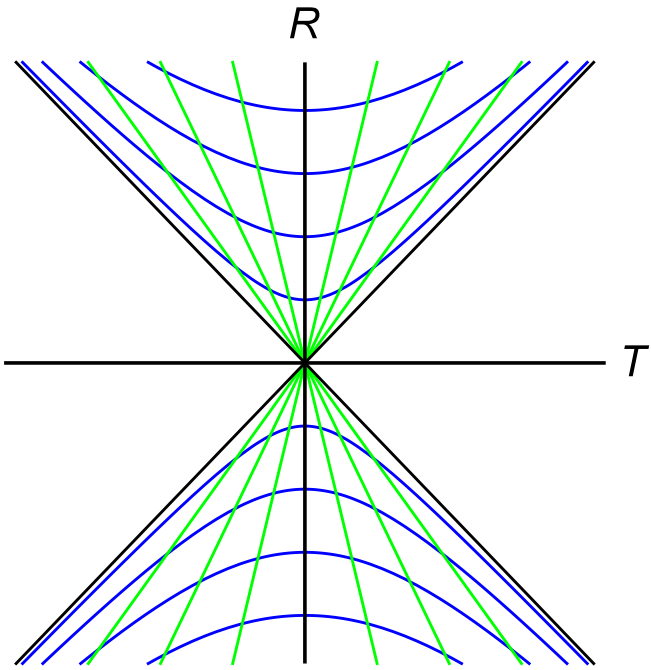} 
		\caption{Hyperbolic coordinates of Minkowski space. Green lines are $r=\text{constant}$, and blue lines are $t=\text{constant}$.}
		\label{hyper}
\end{figure}

Let us start from the flat spherical coordinates $(T,R,\theta,\phi)$ with the metric 
\be
ds^2=-dT^2+dR^2+R^2d\Omega^2.
\ee
We now define the hyperbolic coordinates $(t,r)$ from the relations
\be 
R=\sqrt{-k}\,tr\ ,\quad
T^2-R^2=t^2\ ,\quad
k<0.
\ee
We leave the angular coordinates $(\theta,\phi)$ as they are.

It is straightforward to show that in the hyperbolic coordinates, the metric becomes
\be\lb{hyper_met}
ds^2=-dt+a(t)^2\left(\frac{dr^2}{1-kr^2}+r^2d\Omega\right),\\
\ee
where
\be\lb{hyper_a}
a(t)=\sqrt{-k}\,t.
\ee

Minkowski space can therefore be described by FLRW coordinates for any $k\leq0$. The scale factor of these coordinates will be
\be
a(t) = \left\{
\begin{array}{ll}
	\text{constant} \quad &\text{for } k=0,\\ \\
	\sqrt{-k}\,(t-t_0) \quad &\text{for } k<0,
\end{array}
\right. 
\ee
where $t_0$ is an arbitrary time translation.

For both $k=0$ and $k<0$, we find that that
\be\lb{hyper_fried}
H^2=-\frac{k}{a^2}.
\ee

Given a FLRW metric together with $k=0$ and $H^2=0$, it is always possible to do a coordinate transformation to some other FLRW coordinates with 
\be
\left(H'\right)^2=-\frac{k'}{\left(a'\right)^2},
\ee
with $k'\leq0$.

\end{document}